\documentclass{elsart1p}
 \def\BA{\begin{eqnarray}}
 \def\BE{\begin{equation}}
 \def\BF{\begin{figure}[htb]}
 \def\BT{\begin{table}[htb]}
 \def\EA{\end{eqnarray}}
 \def\EE{\end{equation}}
 \def\EF{\end{figure}}
 \def\ET{\end{table}}

 \def\lsim{\mathrel{\rlap{\lower4pt\hbox{\hskip1pt$\sim$}}
     \raise1pt\hbox{$<$}}}         
 \def\gsim{\mathrel{\rlap{\lower4pt\hbox{\hskip1pt$\sim$}}
     \raise1pt\hbox{$>$}}}         

\newcommand{\AmS}{{\protect\the\textfont2
  A\kern-.1667em\lower.5ex\hbox{M}\kern-.125emS}}

\usepackage{graphicx}
\usepackage{amssymb}
\begin{document}
\begin{frontmatter}
%
%
%
%
%
\vspace*{-1.8cm}
\title{Direct Photon Production in Proton-Nucleus and Nucleus-Nucleus Collisions}
%
%

\author[1]{J. Cepila}
\author[1,2]{J. Nemchik}

\address[1]{
Czech Technical University in Prague, FNSPE,
B\v rehov\' a 7, 11519 Prague, Czech Republic
}
 
\address[2]{
Institute of Experimental Physics SAS,
Watsonova 47, 04001 Ko\v sice, Slovakia
}

\ead{jan.cepila@fjfi.cvut.cz}
\ead{nemchik@saske.sk}

\begin{abstract}

Prompt photons produced in a hard reaction are not
accompanied with any final state interaction, either energy loss or absorption.
Therefore, besides the
Cronin enhancement at medium transverse momenta $p_T$
and small isotopic corrections at larger $p_T$,
one should not expect any nuclear effects.
However, data from PHENIX experiment exhibits a
significant large-$p_T$ suppression in central
$d+Au$ and $Au+Au$ collisions that cannot be accompanied
by coherent phenomena.
We demonstrate that such an unexpected result is subject
to the energy sharing problem near
the kinematic limit and is universally induced
by multiple initial state interactions.
We describe production of photons in the color dipole
approach and find a good agreement with
available data in $p+p$ collisions.
Besides explanation of large-$p_T$ nuclear suppression at
RHIC we present for the first time predictions for expected
nuclear effects also in the LHC energy range at different
rapidities. We include and analyze also a contribution
of gluon shadowing as a leading twist shadowing correction
modifying nuclear effects 
at small and medium $p_T$.

\end{abstract}

\vspace*{-0.7cm}
\begin{keyword}
direct photons \sep
nuclear suppression \sep
gluon shadowing

%

\PACS
13.85.Qk \sep
24.85.+p \sep
25.75.-q \sep
25.75.Cj

\end{keyword}
\end{frontmatter}

\vspace*{-0.6cm}
\section{Introduction}
\label{}
\vspace*{-0.3cm}

If a particle with mass $M$ and transverse momentum
$p_T$ is produced in a hard reaction then the corresponding   
values of Bjorken variable in the beam
and the target are
$x_{1,2} = \sqrt{M^2 + p_T^2}\,e^{\pm y}/\sqrt{s}$.
Thus, forward rapidity region $y > 0$ allows
to study already at RHIC coherence phenomena (shadowing), 
which are expected
to suppress particle yields.

Observed suppression at large $y$
at RHIC \cite{brahms}
should be
interpreted carefully.
Similar suppression is observed 
for any reaction studied so far
at any energy.
Namely, all fixed target experiments
have too low
energy for the onset of coherence effects.
The rise of suppression with $y$
shows the same pattern as observed
at RHIC.

This universality of suppression favors
another mechanism which
was proposed in \cite{knpsj-05} and is based on
energy conservation effects in initial state parton rescatterings.
%
%
%
As a result the effective projectile
parton distribution correlates with the nuclear target
\cite{knpsj-05,prepar1} and can be expressed in term
of the suppression factor, $S(x)\sim 1-x$ \cite{knpsj-05},
\vspace*{-0.2cm}
 \BA
f^{(A)}_{q/N}\bigl (x,
Q^2,{\vec b}\bigr
) =
C\,f_{q/N}\bigl (x,
Q^2\bigr
)\,
exp\biggl[ -[1 - S(x)]\,\sigma_{eff}T_A({\vec b})\biggr]\, ,
\label{10}
 \EA
%
%
where $T_A({\vec b})$ is the nuclear thickness function defined
at impact parameter ${\vec b}$,
$\sigma_{eff} = 20\,$mb \cite{knpsj-05} and
the normalization
factor $C$ is fixed by the Gottfried sum rule.

In this paper we study a production of
direct photons on nuclear targets.
Photons produced in a hard reaction
have no final state interactions and so
no nuclear effects are expected at large $p_T$.
However, we show that large-$p_T$ photons
are universally suppressed by energy deficit in multiple interactions
Eq.~(\ref{10})
since the kinematic limit can be approached increasing
$p_T$ at fixed $y$.
We study also
a rise of this suppression with $y$
in the RHIC and LHC kinematic regions.
\vspace*{-0.3cm}

%
\section{The color dipole approach\label{dipole}}
%
\vspace*{-0.3cm}

The process of direct photon production 
in the target rest frame can
be treated as radiation of a real photon  
by a projectile quark.
The $p_T$ distribution of photon
bremsstrahlung in quark-nucleon interactions
reads \cite{kst1}:
\vspace*{-0.2cm}
%
 \BA
\hspace*{-0.9cm}
\frac{d\sigma(qN\rightarrow \gamma\,X)}{d(ln\,\alpha)\,d^2p_T}
=
\frac{1}{(2\pi)^2}\,
\sum\limits_{in,f}\,
\int d^2 r_1\,d^2 r_2\,
e^{i\vec{p}_T\cdot(\vec{r}_1 - \vec{r}_2)}
\Phi_{\gamma q}^{*~T}(\alpha,\vec{r}_1)
\Phi_{\gamma q}^{T}(\alpha,\vec{r}_2)\,
\Sigma(\alpha,r_1,r_2)
\label{20}
 \EA
%
where
$\Sigma(\alpha,r_1,r_2) = \bigl \{\sigma_{\bar qq}(\alpha r_1)
+ \sigma_{\bar qq}(\alpha r_2)
- \sigma_{\bar qq}(\alpha |\vec{r}_1 - \vec{r}_2|)
\bigr \}/2$,
$\alpha = p^+_{\gamma}/p^+_q$ and
the light-cone (LC) wave functions of the projectile
$q+\gamma$ fluctuation $\Phi_{\gamma q}(\alpha,\vec{r})$ are
presented in \cite{kst1}.
Feynman variable is given as $x_F = x_1 - x_2$ and
in the target rest frame $x_1 = p^+_{\gamma}/p^+_p$.
For the dipole cross section $\sigma_{\bar qq}(\alpha r)$ in Eq.~(\ref{20})
we used GBW \cite{kmw-06}
parametrization. 
The hadron cross section is given convolving the   
parton cross section, Eq.~(\ref{20}), with
the corresponding parton distribution functions (PDFs)
$f_{q}$ and $f_{\bar{q}}$
\cite{kst1},
%
\vspace*{-0.2cm}
 \BA
\hspace*{-.9cm}
\frac{d\sigma(pp\rightarrow \gamma X)}{dx_F\,d^2p_T}
=
\frac{x_1}{x_1 + x_2}
\int_{x_1}^{1}
\frac{d\alpha}{\alpha^2} 
\sum_q Z_q^2
\biggl\{
f_{q}\bigl (\frac{x_1}{\alpha},Q^2\bigr )
+ f_{\bar{q}}\bigl (\frac{x_1}{\alpha},Q^2\bigr ) 
\biggr\}
\frac{d\sigma(qN\to\gamma X)}{d(ln\,\alpha)\,d^2p_T} ,
\label{30}
 \EA
%
where
$Z_q$ is the fractional quark charge,
PDFs $f_q$ and $f_{\bar q}$ are used
with the lowest order parametrization
from \cite{grv98} at the scale
$Q^2 = p_T^2$.  

Assuming production of direct photons on nuclear targets
the onset of coherence effets
is controlled by the
coherence length,
%
$l_c
= 2E_q\,\alpha(1-\alpha) /
(\alpha^2\,m_q^2 + p_T^2)
$,
%
where $E_q = x_q s/2m_N$ and $m_q$ is the energy and mass
of the projectile quark.
The fraction of the proton momentum
$x_q$ carried by the quark is related to
$x_1$ as $\alpha x_q = x_1$.

The condition for the onset of shadowing
is a long coherence
length (LCL), $l_c\gsim R_A$, where $R_A$ is the nuclear radius.
Then the color dipole approach allows to incorporate
shadowing effects via a simple eikonalization
of $\sigma_{\bar qq}(x,r)$  
\cite{zkl},
i.e. replacing $\sigma_{\bar qq}(x,r)$  
in Eq.~(\ref{20}) by 
%
$\sigma_{\bar qq}^A(x,r) =
2 \int d^2 b\,\bigl\{1 - \bigl [1 -
\frac{1}{2\,A}\,\sigma_{\bar qq}(x,r)\,T_A(b)
\bigr ]^A\bigr\}$. 
This LCL limit can be safely used in calculations of
nuclear effects in the RHIC and LHC energy regions
especially at forward rapidities. 
Here higher Fock components containing gluons
lead to additional corrections, called gluon shadowing (GS).
The corresponding suppression factor $R_G$
\cite{knst-02} was included in calculations
replacing 
$\sigma_{\bar qq}$
by $R_G\,\sigma_{\bar qq}$ 
in the above 
expression for $\sigma_{\bar qq}^A(x,r)$.

\begin{figure}[htb]
\vspace*{-0.20cm}

\includegraphics[height=4.6cm,width=6.6cm]{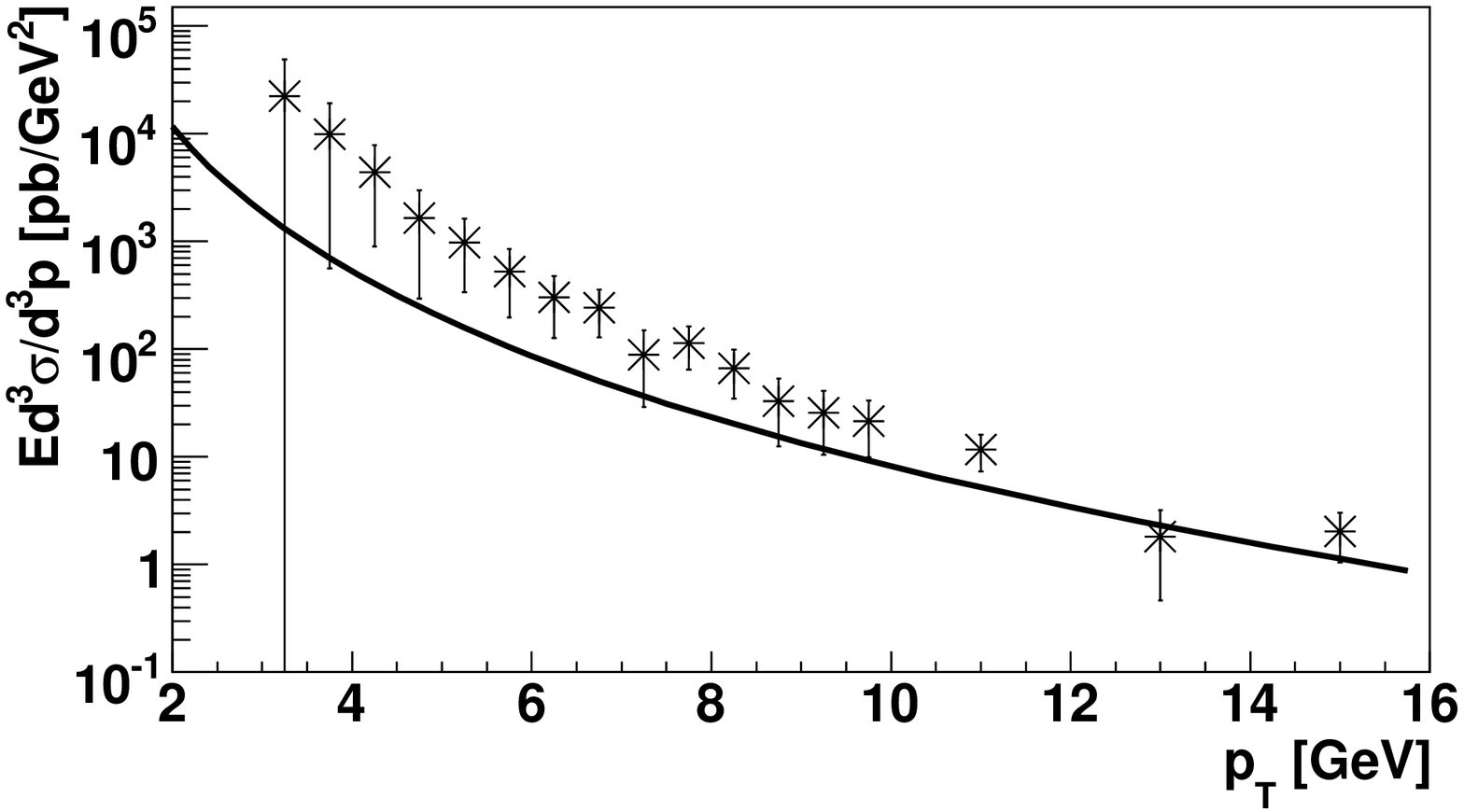}
\hspace*{ 0.1cm}
\includegraphics[height=4.6cm,width=6.6cm]{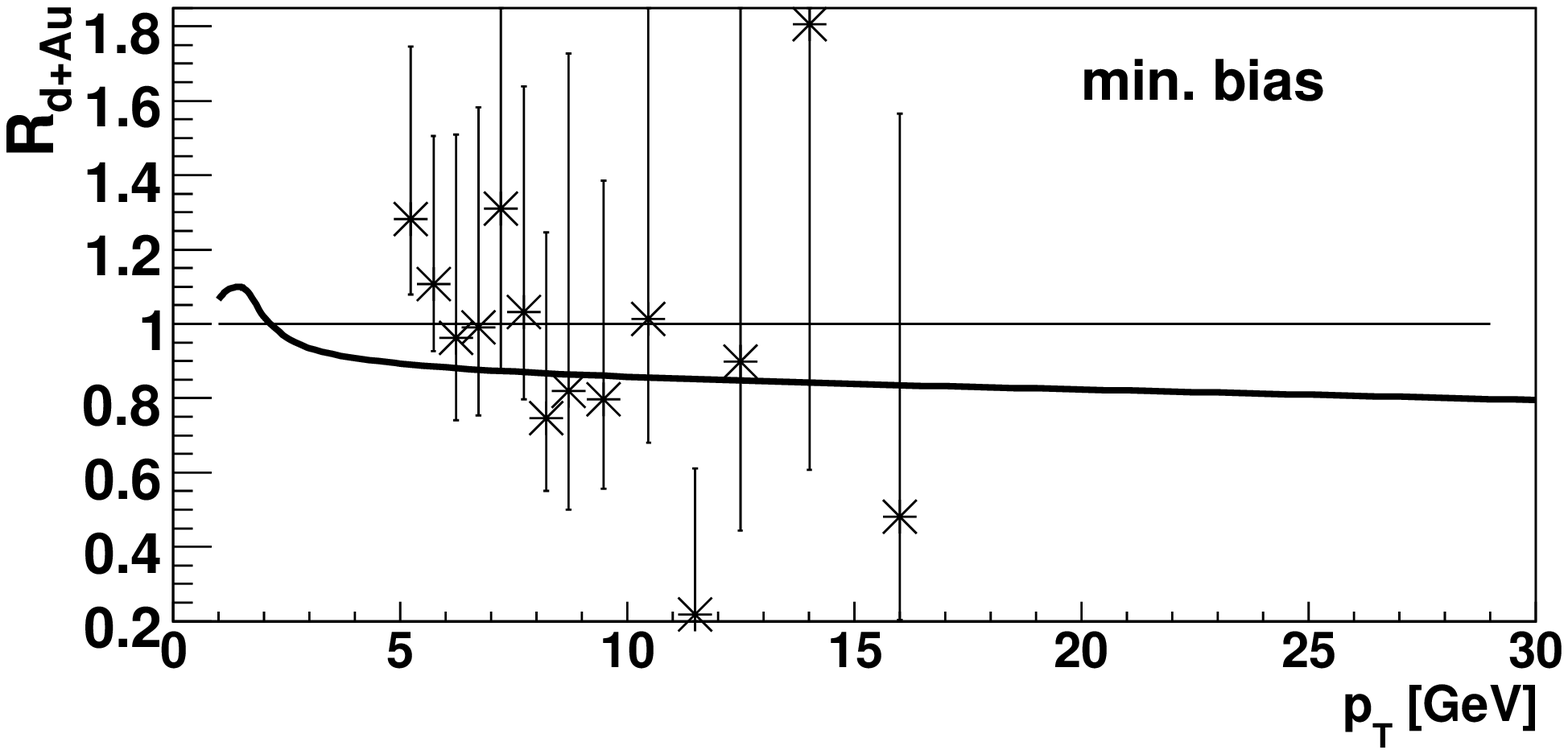}
\begin{center}

\vspace*{-.3cm}   
\caption
{\small
(Left)
Invariant cross section for direct photon production
in $p+p$ collisions at $y = 0$ as a function of $p_T$ vs. data from PHENIX
experiment \cite{dp-phenix-pp}. 
(Right)
Ratio of the cross sections in $d+Au$ to 
$p+p$ collisions $R_{d+Au}(p_T)$ at $\sqrt{s_{NN}} = 200$\,GeV vs. 
preliminary data
from PHENIX experiment \cite{dp-phenix-dAu}.
}
\label{fig1}
\end{center}    
\vspace*{-0.50cm}

\end{figure}

\begin{figure}[htb]

\includegraphics[scale=0.67]{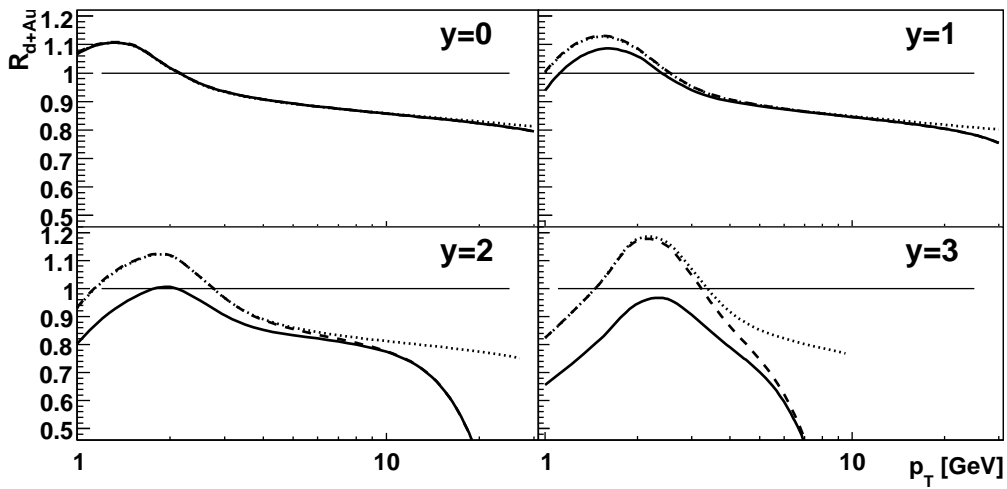}
\begin{center}    

\vspace*{-.2cm}   
\caption
{\small
Ratio of the cross sections in $d+Au$ to $p+p$ collisions 
$R_{d+Au}(p_T)$ at $\sqrt{s_{NN}} = 200$\,GeV and
at different fixed values of $y = $ 0, 1, 2 and 3.
Dotted lines represent calculations without corrections for energy
conservation and GS. Dashed lines additionally include
corrections for energy deficit Eq.~(\ref{10}) and solid lines also GS. 
}
\label{fig2}
\end{center}    
\vspace*{-0.70cm}

\end{figure}

\vspace*{-0.3cm}
%
\section{Predictions for nuclear effects}
%
\vspace*{-0.3cm}

We start with production of
direct photons in $p+p$ collisions.
The left panel of Fig.~\ref{fig1} shows
model calculations based on Eq.~(\ref{30}) using
GRV98 PDFs \cite{grv98} and demonstrates so
a reasonable agreement
with data from PHENIX experiment \cite{dp-phenix-pp}.
Another test of the model is a comparison with 
PHENIX data \cite{dp-phenix-dAu} obtained in $d+Au$ collisions
as is depicted in
the right panel of Fig.~\ref{fig1}. Besides isotopic
effects giving a value
$R_{d+Au}\sim 0.83$ at large $p_T$,
we predict also an additional suppression
coming from corrections for energy conservation Eq.~(\ref{10}).

\begin{figure}[htb]
\hspace*{4.65cm}
\vspace*{-.5cm}

\includegraphics[scale=0.67]{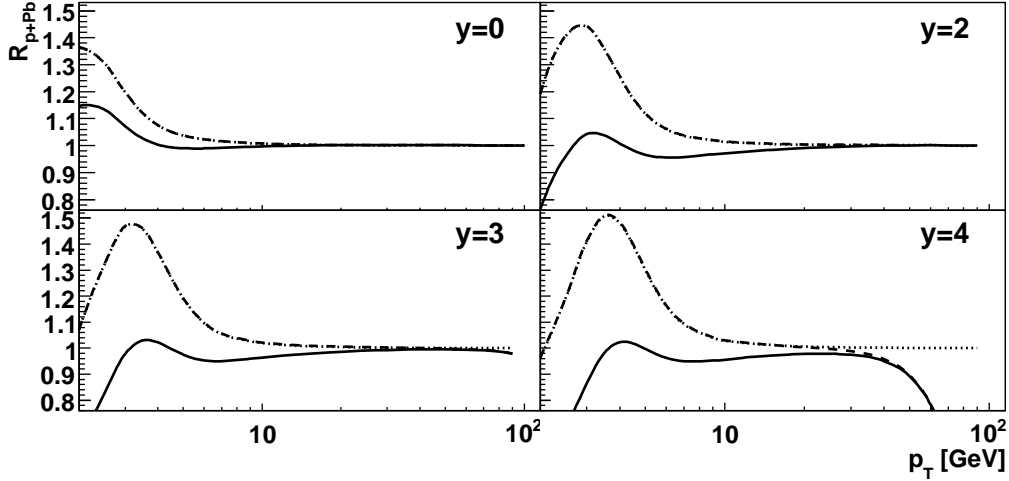}
\begin{center}

\vspace*{-.2cm}   
\caption
{\small
The same as Fig.~\ref{fig2} but for the ratio $R_{p+Pb}(p_T)$
at $\sqrt{s_{NN}} = 5.5$\,TeV and at fixed $y =$0, 2, 3 \& 4. 
}
\label{fig3}
\end{center}    
\vspace*{-0.50cm}

\end{figure}

Since
one can approach the kinematic limit increasing
$p_T$ we present predictions 
for nuclear effects at several fixed $y$
as $p_T$ dependence
of the nuclear modification factor $R_{d+Au}$
at RHIC energy
depicted in Fig.~\ref{fig2} and $R_{p+Pb}$ at LHC energy
depicted in Fig.~\ref{fig3}.
All these Figs. clearly demonstrate a dominance of GS 
at small and medium $p_T$ and energy
conservation effects Eq.~(\ref{10}) at large $p_T$.
Both effects rise rapidly with $y$.
Note that unexpected large-$p_T$ suppression violating
so QCD factorization can be tested 
in the future
by the new data from RHIC and LHC experiments 
especially at forward rapidities.

%
%
%
%

 
The same mechanism
allows to explain also large-$p_T$ suppression of photons
produced in $Au+Au$ collisions
at the energies $\sqrt{s_{NN}} = $200 and 62 GeV in accordance
with data from PHENIX experiment \cite{dp-phenix-AuAu}.
Corresponding results can be found in \cite{prepar1}.
Large error bars of the data do not allow to provide
a definite confirmation for the predicted suppression.

\vspace*{-0.4cm}
%
%
\section{Summary}
%
%
\vspace*{-0.4cm}

Using the color dipole approach
we study production of direct photons in collisions
on nuclear targets.
We demonstrate that at fixed rapidities
effects of coherence (GS) dominate at small and medium $p_T$
whereas corrections for energy conservation 
Eq.~(\ref{10})
are important at larger $p_T$. Both effects 
cause a suppression 
and rise rapidly with rapidity. 

First we test this approach in the RHIC kinematic region
demonstrating a good agreement with PHENIX data
in $p+p$ and $d+Au$ collisons at mid rapidities (see
Fig.~\ref{fig1}).

Then we present predictions for
$p_T$ behavior of
nuclear effects  at different fixed rapidities
in the RHIC and LHC kinematic regions.
Since photons have no final state
interactions, no suppression is expected at large $p_T$.
However, we specify for the first time the kinematic regions
at RHIC and LHC where one
can expect and study in the future a rather strong $p_T$-suppression,
which is caused by energy sharing problem Eq.~(\ref{10}).

The same mechanism explains well also
a strong suppression at large $p_T$ observed
in $Au+Au$ collisions at RHIC in accordance with
data from PHENIX experiment.

\vspace*{-0.3cm}
\ack
\vspace*{-0.3cm}
This work was supported 
by the
Slovak Funding Agency, Grant 2/0092/10 and by Grants VZ M\v SMT
6840770039 and LC 07048 (Ministry of Education of the Czech Rep.).

\end{document}